\documentstyle[amssymb,amstex,preprint,aps]{revtex}

\newtheorem{theorem}{Theorem}
\newtheorem{acknowledgement}[theorem]{Acknowledgement}

\begin{document}
\title{Gaussian, Mean Field and Variational Approximation: the Equivalence}
\author{E. Prodan}
\address{The University of Houston, Dept. of Physics, 4800 Calhoun,\\
Houston, TX 77204-5506}
\maketitle

\begin{abstract}
We show the equivalence between the three approximation schemes for
self-interacting (1+1)-D scalar field theories. Based on rigorous results of
[1, 2], we are able to prove that the Gaussian approximation is very precise
for certain limits of coupling constants. The $\lambda \phi ^{4}+\sigma \phi
^{2}$ model will be used as a concrete application.
\end{abstract}

\section{Introduction}

\noindent We start by clarifying the terms which appear in the title. For a
given theory of self-interacting quantum fields, by Gaussian approximation
we mean the Gaussian part of the interacting measure. By mean field
approximation we understand the leading term of the expansion given in $%
\left[ \text{1, 2}\right] $ and by variational approximation we understand
the old variational technique performed with some particular trial states $%
\left[ \text{3-5}\right] $.

\noindent The novelty of this paper is the method of extracting the Gaussian
peace of the interacting measure. Also we hope that the equivalence between
the three methods will lead to a better understanding of the
self-interacting field theories. The expansion around the mean field (MF)
approximation have been successfully used in $\left[ \text{1, 2}\right] $ to 
$\lambda ^{-2}\sum_{n=1}^{N}\alpha _{n}\left( \lambda \phi \right) ^{2n}$
interactions, for small $\lambda $. The key role of this approximation is
that, after the translation to $\phi -\phi _{MF}$, it brings all the
coupling constants near to zero where the cluster expansion $\left[ \text{6}%
\right] $ can be applied. The results show that the mean field approximation
incorporates the nonanalytic part of the Schwinger functions. Our Gaussian
approximation will play the same role. For $\lambda \rightarrow 0$, $\infty $%
, it provides a transformation of the field which brings all the coupling
constants near to zero. Actually, for $\lambda \rightarrow 0$, the Gaussian
approximation reduces to the mean field approximation. A similar procedure
have been used by Glimm et al in $\left[ \text{7}\right] $ to prove the
equivalence between the model :$\lambda \phi ^{4}+\sigma \phi ^{2}$:, with
large $\lambda $, and the model :$\lambda \phi ^{4}+\sigma \phi ^{2}$: with
small $\lambda $ and negative $\sigma $. However, the variational
approximation scheme (especially in the form of [5]) became very popular
among physicists even thought that there were no estimates of the errors
related to this scheme. The equivalence of the three methods allows us to
use the exact results of [1, 2] to estimate these errors.

\section{Outline of the strategy}

\noindent We consider in this paper $\left( \text{1+1}\right) $-D scalar
fields with self-interactions. To build such theories, one can follow the
strategy presented in [8]. Consider first a space cut-off interaction: 
\begin{equation}
U\left( s\right) =\int_{\left| x\right| <s}d^{2}x:V\left( \phi \left(
x\right) \right) :_{m_{0}}\text{,}
\end{equation}
where the normal ordering is with respect to the vacuum of the free field of
mass $m_{0}$. The next step is the investigation of the cut-off interacting
measure: 
\begin{equation}
d\mu _{s}=%
{\displaystyle{e^{-U\left( s\right) +%
{\textstyle{m_{0}^{2} \over 2}}\int_{\left| x\right| <s}:\left( \phi \left( x\right) -\xi \right) ^{2}:_{m_{0}}}d\mu _{0,\xi } \over \int e^{-U\left( s\right) +%
{\textstyle{m_{0}^{2} \over 2}}\int_{\left| x\right| <s}:\left( \phi \left( x\right) -\xi \right) ^{2}:_{m_{0}}}d\mu _{0,\xi }}}%
=%
{\displaystyle{e^{-U^{\prime }\left( s\right) }d\mu _{0,\xi } \over \int e^{-U^{\prime }\left( s\right) }d\mu _{0,\xi }}}%
\text{,}  \label{fa}
\end{equation}
where $\mu _{0,\xi }$ is the Gaussian measure corresponding to the
covariance $C_{m_{0}}=%
{\displaystyle{1 \over -\Delta +m_{0}^{2}}}%
$ and mean $\xi $. Note that the additional term to the potential $U$
cancels the mass and the mean of the measure $\mu _{0,\xi }$. The last step
is the investigation of the (thermodynamic) limit of this measure: $%
\lim\limits_{s\rightarrow \infty }\mu _{s}$. In the case when the limit is
well defined, one can build the physical Minkowsky field by using one of the
well known reconstruction methods. The hardest part of such a program is
obvious the thermodynamic limit. Our strategy of pursuing this program will
be as follow. The investigation of $\mu _{s}$ is equivalent with the
investigation of the moments (Schwinger functions): 
\begin{equation}
\left\langle \phi \left( x_{1}\right) ...\phi \left( x_{n}\right)
\right\rangle _{s}=\int \phi \left( x_{1}\right) ...\phi \left( x_{n}\right)
d\mu _{s}\text{.}  \label{sch}
\end{equation}
The thermodynamic limit can be studied by considering the dynamical system: 
\begin{equation}
{\displaystyle{d \over ds}}%
\left( 
\begin{array}{c}
\left\langle \phi \left( x_{1}\right) \right\rangle _{s} \\ 
\left\langle \phi \left( x_{1}\right) \phi \left( x_{2}\right) \right\rangle
_{s} \\ 
...
\end{array}
\right) =\vec{X}\text{,}  \label{di}
\end{equation}
where $\vec{X}$ is an infinite vector of whom expression can be deduced by
simply taking the derivative of the right part of \ref{sch} in respect to $s$%
. The thermodynamic limit is reduced now to the study of the stable fixed
points of the dynamical system \ref{di}: if the initial conditions for this
dynamical system lie in the basin of attraction of a stable fixed point then
we found the thermodynamic limit. We will show that the initial conditions
can be modified by changing the boundary conditions of the field. Actually,
in the Gaussian approximation we can choose the initial conditions to
superimpose over the stable fixed points. This will lead to a set of
self-consistency equations which will provide us an approximative value for
the expectation value and the mass of the field. For interactions of the
type: $\lambda ^{-2}P\left( \lambda \phi \right) $, the solution of this set
of equations reveals two things: in the limit $\lambda \rightarrow 0$ we
recover what Glimm et al call the mean field approximation. In the limit $%
\lambda \rightarrow \infty $ (strong couplings), the mass increases much
faster than $\lambda $ so by a rescaling of the field we are again in the
small coupling constants domain. To show the equivalence between Gaussian
and Variational approximations, we compute the vacuum energy in Gaussian
approximation and show that it coincides with the expression given in [5].

\section{The Gaussian approximation}

\noindent Let us start by specifying our notations. In general, by $:\cdot
:_{\mu }$ we denote the normal ordering with respect to some measure $\mu $.
If the measure corresponds to a covariance $C$ and mean $\xi $, we denote it
by $:\cdot :_{C,\xi }$ and further, if the covariance corresponds to some
mass, $m$, then we use $:\cdot :_{m,\xi }$. Instead of $:\cdot :_{\mu _{s}}$
we will use $:\cdot :_{s}$. We denote the expectation values with respect to
some measure $\mu $ by $\left\langle \cdot \right\rangle _{\mu }$ and we
apply the same shorthands as for the normal ordering.

\subsection{The dynamics of the expectation values}

\noindent Starting from: 
\begin{equation}
\left\langle \phi \left( x_{1}\right) ...\phi \left( x_{n}\right)
\right\rangle _{s}=%
{\displaystyle{\int \phi \left( x_{1}\right) ...\phi \left( x_{n}\right) e^{-U^{\prime }\left( s\right) }d\mu _{0} \over \int e^{-U^{\prime }\left( s\right) }d\mu _{0}}}%
\text{,}
\end{equation}
then: 
\begin{equation}
{\displaystyle{d \over ds}}%
\left\langle \phi \left( x_{1}\right) ...\phi \left( x_{n}\right)
\right\rangle _{s}=\left\langle \int_{\left| x\right| =s}dx\ \left[
\left\langle V^{\prime }\left( \phi \left( x\right) \right) \right\rangle
_{s}-V^{\prime }\left( \phi \left( x\right) \right) \right] \phi \left(
x_{1}\right) ...\phi \left( x_{n}\right) \right\rangle _{s}
\end{equation}
or, shortly: 
\begin{equation}
{\displaystyle{d \over ds}}%
\left\langle \phi \left( x_{1}\right) ...\phi \left( x_{n}\right)
\right\rangle _{s}=\left\langle \left[ \left\langle U^{\prime }\left( \delta
_{\Gamma }\right) \right\rangle _{s}-U^{\prime }\left( \delta _{\Gamma
}\right) \right] \phi \left( x_{1}\right) ...\phi \left( x_{n}\right)
\right\rangle _{s},  \label{din}
\end{equation}
where $\Gamma $ is the curve $\left| x\right| =s$ and $\delta $ is the Dirac
delta function. As is stated now, the above equation has the initial
condition: 
\begin{equation}
\left\langle \phi \left( x_{1}\right) ...\phi \left( x_{n}\right)
\right\rangle _{s=0}=\left\langle \phi \left( x_{1}\right) ...\phi \left(
x_{n}\right) \right\rangle _{m_{0},\xi }\text{,}
\end{equation}
in particular: $\left\langle \phi \left( x\right) \right\rangle _{s=0}=\xi $
and $\left\langle :\phi \left( x\right) :_{s=0}:\phi \left( y\right)
:_{s=0}\right\rangle =C_{m_{0}}\left( x-y\right) $.

\subsection{The change of covariance}

\noindent We can allow us a little more liberty in choosing the initial
conditions. If one uses the Gaussian perturbation identity \cite{6}: 
\begin{equation}
d\mu _{\left( C^{-1}+v\right) ^{-1}}=Z^{-1}\exp \left( -:V:_{C}\right) d\mu
_{C}\text{,}
\end{equation}
where $:V:_{C}=%
{\displaystyle{1 \over 2}}%
\int v\left( x\right) :\phi \left( x\right) ^{2}:_{C}$ and $Z\equiv \int
\exp \left( -:V:_{C}\right) d\phi _{C}$, it follows that, up to a change of
boundary conditions: 
\begin{equation}
d\mu _{s}=%
{\displaystyle{e^{-U^{\prime }\left( s\right) }d\mu _{0} \over \int e^{-U^{\prime }\left( s\right) }d\mu _{0}}}%
=%
{\displaystyle{e^{-\tilde{U}\left( s\right) }d\mu _{C_{m},\xi } \over \int e^{-\tilde{U}\left( s\right) }d\mu _{C_{m},\xi }}}%
\text{,}  \label{tr}
\end{equation}
where: 
\begin{equation}
\tilde{U}\left( s\right) \equiv \int_{\left| x\right| <s}d^{2}x\,\tilde{V}%
\left( \phi \left( x\right) \right) =\int_{\left| x\right| <s}d^{2}x\
\left\{ :V\left( \phi \left( x\right) \right) :_{m_{0}}-%
{\displaystyle{m^{2} \over 2}}%
:\left( \phi \left( x\right) -\xi \right) ^{2}:_{m_{0}}\right\}  \label{npot}
\end{equation}
and $d\mu _{C_{m},\xi }$ is the Gaussian measure corresponding to the
covariance $C_{m}=%
{\displaystyle{1 \over -\Delta +m^{2}}}%
$ and mean $\xi $. Now it is the time to discuss the boundary conditions.
One can see that from very beginning some boundary conditions were imposed.
For example in \ref{fa}, the additional term cancels the mass and mean of
the measure $d\mu _{0,\xi }$ only for $\left| x\right| <s$. This cut-off is
equivalent with adding an interaction in the region $\left| x\right| >s$. In
the limit $s\rightarrow \infty $, it provides us the external conditions.
One can read [9] for a discussion of this boundary conditions. The same
remarks for \ref{tr}: the Gaussian perturbation identity was cut off, the
effect being the change of the external conditions. In the light of this
remarks, one can see that all the time when $m$ and $\xi $ are changed, the
boundary conditions are changed or, reciprocally, we can change $m$ and $\xi 
$ by a modification of the boundary conditions. As a result, we can control
the initial conditions of the dynamical system \ref{din} by imposing
different boundary conditions: 
\begin{equation}
\left\{ 
\begin{array}{l}
{\displaystyle{d \over ds}}%
\left\langle \phi \left( x_{1}\right) ...\phi \left( x_{n}\right)
\right\rangle _{s}=\left\langle \left[ \left\langle \tilde{U}\left( \delta
_{\Gamma }\right) \right\rangle _{s}-\tilde{U}\left( \delta _{\Gamma
}\right) \right] \phi \left( x_{1}\right) ...\phi \left( x_{n}\right)
\right\rangle _{s} \\ 
\\ 
\left\langle \phi \left( x_{1}\right) ...\phi \left( x_{n}\right)
\right\rangle _{s=0}=\left\langle \phi \left( x_{1}\right) ...\phi \left(
x_{n}\right) \right\rangle _{m,\xi }
\end{array}
\right. \text{.}  \label{fin}
\end{equation}
If the system is multiphasic, then \ref{fin} will have more than one stable
fixed point. The above procedure can be used in the following way to select
a pure phase: enforce the thermodynamic limit to converge to a fixed stable
point by choosing the initial conditions in the basin of attraction of this
point. There is now a nice picture of critical phenomena in terms of the
bifurcation theory. Suppose \ref{fin} has a degenerate fixed point and there
is a bifurcation point where the degeneracy is lifted. Then the critical
values of the parameters are given by the bifurcation point.

\subsection{The Gaussian approximation}

\noindent We consider that $\mu _{s}$ is approximately Gaussian and
symmetric at translations, or one can think that from now we are interested
only in the symmetric Gaussian part (and the mean) of the measure $\mu _{s}$%
. The major property of a Gaussian measure is that the normal ordered powers
are orthogonal: 
\begin{equation}
\left\langle :\phi ^{n}:_{s}:\phi ^{m}:_{s}\right\rangle _{s}=0\text{ \ if \ 
}m\neq n\text{.}
\end{equation}
We will select the following two equations of the dynamical system \ref{din}
which in the Gaussian approximation write: 
\begin{equation}
\left\{ 
\begin{array}{l}
{\displaystyle{d \over ds}}%
\left\langle \phi \left( 0\right) \right\rangle _{s}=-T_{1,s}\left( \tilde{V}%
\left( \phi \left( x\right) \right) \right) \left\langle :\phi \left( \delta
_{\Gamma }\right) :_{s}:\phi \left( 0\right) :_{s}\right\rangle _{s} \\ 
\\ 
{\displaystyle{d \over ds}}%
\left( \left\langle :\phi \left( 0\right) :_{s}^{2}\right\rangle
_{s}-\left\langle :\phi \left( x\right) :_{m_{0}}^{2}\right\rangle
_{m_{0}}\right) =-T_{2,s}\left( \tilde{V}\left( \phi \left( x\right) \right)
\right) \left\langle :\phi ^{2}\left( \delta _{\Gamma }\right) :_{s}:\phi
\left( 0\right) :_{s}^{2}\right\rangle _{s}
\end{array}
\right.  \label{gaus}
\end{equation}
where $T_{n,s}\left( \cdot \right) $ means the coefficient of $:\phi
^{n}:_{s}$ of the given expression. The point $x$ satisfies $\left| x\right|
=s$ and we have used the symmetry of the problem to separate the terms. Note
that the second equation can be written in the above form because $%
\left\langle \tilde{U}\left( \delta _{\Gamma }\right) \right\rangle _{s}-%
\tilde{U}\left( \delta _{\Gamma }\right) $ has no free term and $:\phi
^{2}:_{s}-:\phi :_{s}^{2}$ is a constant. We have taken the combination $%
\left\langle :\phi \left( 0\right) :_{s}^{2}\right\rangle _{s}-\left\langle
:\phi \left( x\right) :_{m_{0}}^{2}\right\rangle _{m_{0}}$ because it is a
finite quantity and for the reason explained in the following. After one
changes the normal ordering in $\tilde{V}\left( \phi \right) $ from $:\cdot
:_{m_{0}}$ to $:\cdot :_{s}$, 
\begin{equation}
\left. 
\begin{array}{l}
:\phi \left( x\right) ^{n}:_{m_{0}}=\lim\limits_{a\rightarrow
0}\sum_{k=0}^{n}C_{n}^{k}\times \\ 
\\ 
\times 
{\displaystyle{\partial ^{n-k} \over \partial a^{n-k}}}%
\left\{ \exp \left( 
{\displaystyle{a^{2} \over 2}}%
\left( \left\langle :\phi \left( x\right) :_{s}^{2}\right\rangle
_{s}-\left\langle :\phi \left( x\right) :_{m_{0}}^{2}\right\rangle
_{m_{0}}\right) +a\left\langle \phi \left( x\right) \right\rangle
_{s}\right) \right\} :\phi \left( x\right) ^{k}:_{s}\text{,}
\end{array}
\right.  \label{che}
\end{equation}
relation which is valid in the Gaussian approximation, one can see that $%
T_{n,s}\left( \tilde{V}\left( \phi \left( x\right) \right) \right) $ is a
function only of the combination $\left\langle :\phi \left( x\right)
:_{s}^{2}\right\rangle _{s}-\left\langle :\phi \left( x\right)
:_{m_{0}}^{2}\right\rangle _{m_{0}}$ and the expectation value $\left\langle
\phi \left( x\right) \right\rangle _{s}$. Now, because we consider also the
symmetric part of $\mu _{s}$ at translations, the above system can be
written in the form: 
\begin{equation}
\left\{ 
\begin{array}{l}
{\displaystyle{d \over ds}}%
\left\langle \phi \left( x\right) \right\rangle _{s}=-T_{1,s}\left( \tilde{V}%
\left( \phi \left( x\right) \right) \right) \left\langle :\phi \left( \delta
_{\Gamma }\right) :_{s}:\phi \left( 0\right) :_{s}\right\rangle _{s} \\ 
\\ 
{\displaystyle{d \over ds}}%
\left( \left\langle :\phi \left( x\right) :_{s}^{2}\right\rangle
_{s}-\left\langle :\phi \left( x\right) :_{m_{0}}^{2}\right\rangle
_{m_{0}}\right) =-T_{2,s}\left( \tilde{V}\left( \phi \left( x\right) \right)
\right) \left\langle :\phi ^{2}\left( \delta _{\Gamma }\right) :_{s}:\phi
\left( 0\right) :_{s}^{2}\right\rangle _{s} \\ 
\\ 
\left\langle \phi \left( x\right) \right\rangle _{s=0}=\xi \text{ , }\left(
\left\langle :\phi \left( x\right) :_{s=0}^{2}\right\rangle
_{s=0}-\left\langle :\phi \left( x\right) :_{m_{0}}^{2}\right\rangle
_{m_{0}}\right) =%
{\displaystyle{1 \over 4\pi }}%
\ln \left( 
{\displaystyle{m_{0}^{2} \over m^{2}}}%
\right) \text{.}
\end{array}
\right.
\end{equation}
Let us rewrite this system in the following notations: 
\begin{equation}
\left\{ 
\begin{array}{l}
X=\left\langle \phi \left( x\right) \right\rangle _{s} \\ 
\\ 
Y=\left\langle :\phi \left( x\right) :_{s}^{2}\right\rangle
_{s}-\left\langle :\phi \left( x\right) :_{m_{0}}^{2}\right\rangle _{m_{0}}%
\text{.}
\end{array}
\right.
\end{equation}
Then: 
\begin{equation}
\left\{ 
\begin{array}{l}
{\displaystyle{dX \over ds}}%
=-T_{1,s}\left( X,Y\right) \left\langle :\phi \left( \delta _{\Gamma
}\right) :_{s}:\phi \left( 0\right) :_{s}\right\rangle _{s} \\ 
\\ 
{\displaystyle{dY \over ds}}%
=-T_{2,s}\left( X,Y\right) \left\langle :\phi ^{2}\left( \delta _{\Gamma
}\right) :_{s}:\phi \left( 0\right) :_{s}^{2}\right\rangle _{s} \\ 
\\ 
X\left( 0\right) =\xi \text{ , }Y\left( 0\right) =%
{\displaystyle{1 \over 4\pi }}%
\ln \left( 
{\displaystyle{m_{0}^{2} \over m^{2}}}%
\right) \text{.}
\end{array}
\right.
\end{equation}
It is easily now to find the fixed points of the dynamical system (more
exactly, the first two coordinates of the fixed points), which are given by
the conditions: 
\begin{equation}
T_{1,s}\left( X,Y\right) =0\text{ , }T_{2,s}\left( X,Y\right) =0\text{.}
\end{equation}
Because we do not have to much information about the basin of attraction of
the fixed points, the best thing we can do is to superimpose the initial
conditions over the stable points. This leads to our self-consistency
equations: 
\begin{equation}
T_{1,s}\left( \xi ,%
{\displaystyle{1 \over 4\pi }}%
\ln \left( 
{\displaystyle{m_{0}^{2} \over m^{2}}}%
\right) \right) =0\text{ , }T_{2,s}\left( \xi ,%
{\displaystyle{1 \over 4\pi }}%
\ln \left( 
{\displaystyle{m_{0}^{2} \over m^{2}}}%
\right) \right) =0\text{,}
\end{equation}
which provide us the Gaussian approximation of the expectation value $%
\left\langle \phi \left( x\right) \right\rangle _{s\rightarrow \infty }=\xi $
and the two point Schwinger function $\left\langle :\phi \left( x_{1}\right)
:_{s\rightarrow \infty }:\phi \left( x_{2}\right) :_{s\rightarrow \infty
}\right\rangle _{s\rightarrow \infty }=C_{m}\left( x_{1},x_{2}\right) $.

\section{The equivalence with the Variational Method}

\noindent We calculate the energy of the vacuum state in the Gaussian
approximation. Will follow that our expression is the same with that of [5].
Moreover, our self-consistency equations can be derived by minimizing this
energy with respect to $\xi $ and $m$.

\noindent The term $Z$, which appears in the Gaussian perturbation identity,
was unimportant for the expectation values but will be essential for the
calculus of the vacuum energy. According to \cite{8}, the energy per unit
length can be calculated as: 
\begin{equation}
\varepsilon =-\lim\limits_{s\rightarrow \infty }%
{\displaystyle{\ln \int e^{-U^{\prime }\left( s\right) }d\mu _{0,\xi } \over \pi s^{2}}}%
\text{.}
\end{equation}
After the change of the covariance: 
\begin{equation}
\varepsilon =-\lim\limits_{s\rightarrow \infty }%
{\displaystyle{\ln Z^{-1}\int e^{-\tilde{U}\left( s\right) }d\mu _{C_{m},\xi } \over \pi s^{2}}}%
\text{,}
\end{equation}
with: 
\begin{equation}
Z=\int \exp \left( 
{\displaystyle{m^{2}-m_{0}^{2} \over 2}}%
\int d^{2}x\,:\left( \phi \left( x\right) -\xi \right) ^{2}:_{m_{0}}\right)
d\mu _{0,\xi }\text{.}
\end{equation}
which can be calculated formally \cite{6} as: 
\begin{equation}
Z=\exp \left( 
{\displaystyle{1 \over 2}}%
Tr\left\{ \ln \left( 1+%
{\displaystyle{m^{2}-m_{0}^{2} \over -\Delta +m_{0}^{2}}}%
\right) -%
{\displaystyle{m^{2}-m_{0}^{2} \over -\Delta +m_{0}^{2}}}%
\right\} \right) \text{.}
\end{equation}
In this form, the above quantity is infinite. Taking into account that when
we have performed the change of covariance we have considered the integral $%
\int d^{2}x\,:\left( \phi \left( x\right) -\xi \right) :_{m_{0}}$ only on $%
\left| x\right| <s$, we should do the same thing for $Z$. Considering the
operator: 
\begin{equation}
\hat{K}=\ln \left( 1+%
{\displaystyle{m^{2}-m_{0}^{2} \over -\Delta +m_{0}^{2}}}%
\right) -%
{\displaystyle{m^{2}-m_{0}^{2} \over -\Delta +m_{0}^{2}}}%
\text{,}
\end{equation}
with the kernel: 
\begin{equation}
K\left( x,y\right) =\int d^{2}k\left\{ \ln \left( 1+%
{\displaystyle{1 \over 4\pi ^{2}}}%
{\displaystyle{m^{2}-m_{0}^{2} \over kh2+m_{0}^{2}}}%
\right) -%
{\displaystyle{1 \over 4\pi ^{2}}}%
{\displaystyle{m^{2}-m_{0}^{2} \over k^{2}+m_{0}^{2}}}%
\right\} e^{-ik\left( x-y\right) }\text{,}
\end{equation}
it follows: 
\begin{equation}
Tr\left[ \hat{K}\right] =\int_{\left| x\right| <s}d^{2}x\int d^{2}k\left\{
\ln \left( 1+%
{\displaystyle{1 \over 4\pi ^{2}}}%
{\displaystyle{m^{2}-m_{0}^{2} \over kh2+m_{0}^{2}}}%
\right) -%
{\displaystyle{1 \over 4\pi ^{2}}}%
{\displaystyle{m^{2}-m_{0}^{2} \over k^{2}+m_{0}^{2}}}%
\right\}
\end{equation}
if we restrict the domain of integration at $\left| x\right| <s$. Finally: 
\begin{equation}
Tr\left[ \hat{K}\right] =%
{\displaystyle{\pi s^{2} \over 4\pi }}%
\left( m^{2}-m_{0}^{2}-m^{2}\ln 
{\displaystyle{m^{2} \over m_{0}^{2}}}%
\right) \text{.}
\end{equation}
In the Gaussian approximation, the other term of $\varepsilon $: -$%
{\displaystyle{\ln \left\langle e^{-\tilde{U}\left( s\right) }\right\rangle _{m,\xi } \over \pi s^{2}}}%
$ is given by $T_{0,s}\left( \tilde{V}\right) .$ In consequence: 
\begin{equation}
\varepsilon =T_{0,s}\left( \tilde{V}\right) +%
{\displaystyle{1 \over 8\pi }}%
\left( m^{2}-m_{0}^{2}-m^{2}\ln 
{\displaystyle{m^{2} \over m_{0}^{2}}}%
\right) \text{.}
\end{equation}
The last expression is completely equivalent with that of [5]. Indeed, for
an analytic function, $F$, it follows from \ref{che} that: 
\[
\begin{array}{l}
T_{0,s}\left[ :F\left( \phi \left( x\right) \right) :_{m_{0}}\right]
=\sum_{n=0}^{\infty }%
{\displaystyle{1 \over n!}}%
\left. 
{\displaystyle{\partial ^{n}F\left( x\right)  \over \partial x^{n}}}%
\right| _{x=0}%
{\displaystyle{\partial ^{n} \over \partial a^{n}}}%
\left. \left\{ \exp \left( 
{\displaystyle{a^{2} \over 8\pi }}%
\ln 
{\displaystyle{m_{0}^{2} \over m^{2}}}%
+a\xi \right) \right\} \right| _{a=0} \\ 
\\ 
=\sum_{n=0}^{\infty }\sum_{k=0}^{n}%
{\displaystyle{1 \over \left( n-k\right) !k!}}%
\left. 
{\displaystyle{\partial ^{n}F \over \partial x^{n}}}%
\right| _{x=0}\xi ^{n-k}\left. 
{\displaystyle{\partial ^{k}\exp \left( %
{\displaystyle{a^{2} \over 8\pi }}\ln %
{\displaystyle{m_{0}^{2} \over m^{2}}}\right)  \over \partial a^{k}}}%
\right| _{a=0} \\ 
\\ 
=\sum_{n=0}^{\infty }\sum_{k=0}^{n}%
{\displaystyle{1 \over \left( n-k\right) !k!}}%
\left. 
{\displaystyle{\partial ^{n}F \over \partial x^{n}}}%
\right| _{x=0}\xi ^{n-k}\left. 
{\displaystyle{\partial ^{k} \over \partial a^{k}}}%
\sum_{p=0}^{\infty }%
{\displaystyle{1 \over p!}}%
\left( 
{\displaystyle{a^{2} \over 8\pi }}%
\ln 
{\displaystyle{m_{0}^{2} \over m^{2}}}%
\right) ^{p}\right| _{a=0} \\ 
\\ 
=\sum_{n=0}^{\infty }\sum_{p=0}^{\infty }%
{\displaystyle{1 \over \left( n-2p\right) !\left( 2p\right) !}}%
\left. 
{\displaystyle{\partial ^{n}F \over \partial x^{n}}}%
\right| _{x=0}\xi ^{n-2p}%
{\displaystyle{\left( 2p\right) ! \over p!}}%
\left( 
{\displaystyle{1 \over 8\pi }}%
\ln 
{\displaystyle{m_{0}^{2} \over m^{2}}}%
\right) ^{p} \\ 
\\ 
=\sum_{p=0}^{\infty }%
{\displaystyle{1 \over p!}}%
\left( 
{\displaystyle{1 \over 8\pi }}%
\ln 
{\displaystyle{m_{0}^{2} \over m^{2}}}%
\right) ^{p}\sum_{n=0}^{\infty }%
{\displaystyle{1 \over \left( n-2p\right) !}}%
\left. 
{\displaystyle{\partial ^{n}F \over \partial x^{n}}}%
\right| _{x=0}\xi ^{n-2p}=\left. \exp \left( 
{\displaystyle{1 \over 8\pi }}%
\ln 
{\displaystyle{m_{0}^{2} \over m^{2}}}%
{\displaystyle{d^{2} \over dx^{2}}}%
\right) F\left( x\right) \right| _{x=\xi }\text{.}
\end{array}
\]
With this result we have: 
\begin{equation}
\left. 
\begin{array}{l}
\varepsilon \left( m,\xi \right) =\left. \exp \left( 
{\displaystyle{1 \over 8\pi }}%
\ln 
{\displaystyle{m_{0}^{2} \over m^{2}}}%
{\displaystyle{d^{2} \over dx^{2}}}%
\right) \left( V\left( \xi \right) -%
{\displaystyle{m^{2} \over 2}}%
\left( x-\xi \right) ^{2}\right) \right| _{x=\xi } \\ 
\\ 
+%
{\displaystyle{1 \over 8\pi }}%
\left( m^{2}-m_{0}^{2}-m^{2}\ln 
{\displaystyle{m^{2} \over m_{0}^{2}}}%
\right) =\exp \left( 
{\displaystyle{1 \over 8\pi }}%
\ln 
{\displaystyle{m_{0}^{2} \over m^{2}}}%
{\displaystyle{d^{2} \over d\xi ^{2}}}%
\right) V\left( \xi \right) +%
{\displaystyle{1 \over 8\pi }}%
\left( m^{2}-m_{0}^{2}\right) \text{.}
\end{array}
\right.
\end{equation}
We show in the following that our self-consistency equations can be also
derived by minimizing $\varepsilon $ in respect to $\xi $ and $m$. Indeed: 
\begin{equation}
\left. 
\begin{array}{l}
T_{1,s}\left[ :\tilde{V}\left( \phi \left( x\right) \right) :_{m_{0}}\right]
=\sum_{n=0}^{\infty }%
{\displaystyle{1 \over n!}}%
\left. 
{\displaystyle{\partial ^{n}\tilde{V}\left( x\right)  \over \partial x^{n}}}%
\right| _{x=0}n%
{\displaystyle{\partial ^{n-1} \over \partial a^{n-1}}}%
\left. \left\{ \exp \left( 
{\displaystyle{a^{2} \over 8\pi }}%
\ln 
{\displaystyle{m_{0}^{2} \over m^{2}}}%
+a\xi \right) \right\} \right| _{a=0} \\ 
\\ 
=\sum_{n=0}^{\infty }%
{\displaystyle{1 \over n!}}%
\left. 
{\displaystyle{\partial ^{n}\tilde{V} \over \partial x^{n}}}%
\right| _{x=0}%
{\displaystyle{\partial ^{n} \over \partial a^{n}}}%
\left. \left\{ a\exp \left( 
{\displaystyle{a^{2} \over 8\pi }}%
\ln 
{\displaystyle{m_{0}^{2} \over m^{2}}}%
+a\xi \right) \right\} \right| _{a=0} \\ 
\\ 
=%
{\displaystyle{\partial  \over \partial \xi }}%
\sum_{n=0}^{\infty }%
{\displaystyle{1 \over n!}}%
\left. 
{\displaystyle{\partial ^{n}\tilde{V} \over \partial x^{n}}}%
\right| _{x=0}%
{\displaystyle{\partial ^{n} \over \partial a^{n}}}%
\left. \left\{ \exp \left( 
{\displaystyle{a^{2} \over 8\pi }}%
\ln 
{\displaystyle{m_{0}^{2} \over m^{2}}}%
+a\xi \right) \right\} \right| _{a=0}=%
{\displaystyle{\partial \varepsilon  \over \partial \xi }}%
\text{.}
\end{array}
\right.
\end{equation}
Using the same notation as before, $Y=%
{\textstyle{1 \over 8\pi }}%
\ln 
{\textstyle{m_{0}^{2} \over m^{2}}}%
$, for the second equations we have: 
\begin{equation}
\left. 
\begin{array}{l}
T_{2,s}\left[ :\tilde{V}\left( \phi \left( x\right) \right) :_{m_{0}}\right]
=\sum_{n=0}^{\infty }%
{\displaystyle{1 \over n!}}%
\left. 
{\displaystyle{\partial ^{n}\tilde{V} \over \partial x^{n}}}%
\right| _{x=0}n\left( n-1\right) \left. 
{\displaystyle{\partial ^{n-2} \over \partial a^{n-2}}}%
\left\{ \exp \left( 
{\displaystyle{a^{2} \over 8\pi }}%
\ln 
{\displaystyle{m_{0}^{2} \over m^{2}}}%
+a\xi \right) \right\} \right| _{a=0} \\ 
\\ 
=\sum_{n=0}^{\infty }%
{\displaystyle{1 \over n!}}%
\left. 
{\displaystyle{\partial ^{n}\tilde{V} \over \partial x^{n}}}%
\right| _{x=0}\left. 
{\displaystyle{\partial ^{n} \over \partial a^{n}}}%
\left\{ a^{2}\exp \left( a^{2}Y+a\xi \right) \right\} \right| _{a=0} \\ 
\\ 
=\sum_{n=0}^{\infty }%
{\displaystyle{1 \over n!}}%
\left\{ 
{\displaystyle{\partial  \over \partial Y}}%
\left( \left. 
{\displaystyle{\partial ^{n}\tilde{V} \over \partial x^{n}}}%
\right| _{x=0}\left. 
{\displaystyle{\partial ^{n} \over \partial a^{n}}}%
\left\{ \exp \left( a^{2}Y+a\xi \right) \right\} \right| _{a=0}\right)
\right. \\ 
\\ 
\left. -%
{\displaystyle{\partial  \over \partial Y}}%
\left( \left. 
{\displaystyle{\partial ^{n}\tilde{V} \over \partial x^{n}}}%
\right| _{x=0}\right) \left. 
{\displaystyle{\partial ^{n} \over \partial a^{n}}}%
\left\{ \exp \left( a^{2}Y+a\xi \right) \right\} \right| _{a=0}\right\} \\ 
\\ 
=%
{\displaystyle{\partial  \over \partial Y}}%
\left[ T_{0,s}\left( \tilde{V}\right) \right] +\sum_{n=0}^{2}%
{\displaystyle{1 \over n!}}%
{\displaystyle{\partial  \over \partial Y}}%
\left( \left. 
{\displaystyle{\partial ^{n}\left[ %
{\textstyle{m^{2} \over 2}}\left( x-\xi \right) ^{2}\right]  \over \partial x^{n}}}%
\right| _{x=0}\right) \left. 
{\displaystyle{\partial ^{n} \over \partial a^{n}}}%
\left\{ \exp \left( a^{2}Y+a\xi \right) \right\} \right| _{a=0}\text{.}
\end{array}
\right.  \label{fas}
\end{equation}
Using that: 
\begin{equation}
{\displaystyle{\partial  \over \partial Y}}%
\left[ T_{0,s}\left( \tilde{V}\right) \right] =%
{\displaystyle{\partial  \over \partial Y}}%
\left\{ \varepsilon -m^{2}\left( 
{\displaystyle{1 \over 8\pi }}%
+Y\right) \right\} \,,\,%
{\displaystyle{\partial  \over \partial Y}}%
\left\{ m^{2}\left( 
{\displaystyle{1 \over 8\pi }}%
+Y\right) \right\} =Y%
{\displaystyle{\partial m^{2} \over \partial Y}}%
\end{equation}
and expanding the second term of \ref{fas}, it follows: 
\begin{equation}
T_{2,s}\left[ :\tilde{V}\left( \phi \left( x\right) \right) :_{m_{0}}\right]
=%
{\displaystyle{\partial \varepsilon  \over \partial Y}}%
\text{.}
\end{equation}
We have proven that the Gaussian approximation of the vacuum energy density
is equal with the expression obtained in [5] and that our self-consistency
equations can be also derived by minimizing the energy density with respect
to $\xi $ and $m$. In consequence the two methods are completely equivalent.

\section{The $V\left( \protect\phi \right) =\protect\lambda \protect\phi
^{4}+\protect\sigma \protect\phi ^{2}$ model}

\noindent We apply in this section the Gaussian approximation to this
particular model. The self-consistency equations are: 
\begin{equation}
\left\{ 
\begin{array}{l}
T_{1,s}\left[ :\tilde{V}\left( \phi \right) :_{m_{0}}\right] =%
{\displaystyle{\xi  \over \pi }}%
\left( 3\lambda \ln 
{\displaystyle{m_{0}^{2} \over m^{2}}}%
+4\pi \lambda \xi ^{2}+2\pi \sigma \right) =0 \\ 
\\ 
T_{2,s}\left[ :\tilde{V}\left( \phi \right) :_{m_{0}}\right] =%
{\displaystyle{3\lambda  \over \pi }}%
\left( \ln 
{\displaystyle{m_{0}^{2} \over m^{2}}}%
+4\pi \xi ^{2}\right) +\sigma -%
{\displaystyle{m^{2} \over 2}}%
=0\text{.}
\end{array}
\right.
\end{equation}
Note that the first equation always has the solution $\xi =0$. Let us
analyze first this situation. The second equation leads to: 
\begin{equation}
m^{2}=%
{\displaystyle{3\lambda  \over \pi }}%
W_{0}\left( 
{\displaystyle{m_{0}^{2}\pi  \over 3\lambda }}%
\exp \left( 
{\displaystyle{2\pi \sigma  \over 3\lambda }}%
\right) \right) \text{,}  \label{ze}
\end{equation}
where $W_{0}$ is the Lambert $W$ function of rank zero. For $\sigma >0,$ we
can introduce the classical mass: $m_{c}^{2}=V^{\prime \prime }\left(
0\right) =2\sigma $. One can see that, if we choose $m_{0}=m_{c}$, then $%
m_{c}$ is a solution of the above equation. For $\sigma <0$, the solution $%
\xi =0$ becomes unstable. The other solutions must satisfy: 
\begin{equation}
\left\{ 
\begin{array}{l}
m^{2}=m_{0}^{2}\exp \left( 
{\displaystyle{2\pi  \over 3\lambda }}%
\left( 2\lambda \xi ^{2}+\sigma \right) \right) \\ 
\\ 
m^{2}=8\lambda \xi ^{2}\text{.}
\end{array}
\right.  \label{syy}
\end{equation}
For $\sigma <0$ we can define the mean field approximation: $\left( \xi
_{c}\right) ^{2}=-%
{\displaystyle{\sigma  \over 2\lambda }}%
$ and $m_{c}^{2}=V^{\prime \prime }\left( \xi _{c}\right) =8\lambda \xi
_{c}^{2}=-4\sigma $. One can see that, if we start with $m_{0}=m_{c}$, then
the mean field approximation is a solution of our self-consistency
equations. This is true for more general interactions and it shows the link
between the Gaussian and mean field approximations. Fortunately this is not
the only solution. The general solution of \ref{syy} (for $m_{0}=m_{c}$) is: 
\begin{equation}
\xi ^{2}=%
{\displaystyle{-3 \over 4\pi }}%
W\left( 
{\displaystyle{2\pi \sigma  \over 3\lambda }}%
\exp \left( 
{\displaystyle{2\pi \sigma  \over 3\lambda }}%
\right) \right) \text{,}
\end{equation}
where $W$ can be the Lambert function of rank $0$ or $-1$. It follows that: 
\begin{equation}
\xi ^{2}=\left\{ 
\begin{array}{c}
{\displaystyle{-3 \over 4\pi }}%
W_{-1}\left( 
{\displaystyle{2\pi \sigma  \over 3\lambda }}%
\exp \left( 
{\displaystyle{2\pi \sigma  \over 3\lambda }}%
\right) \right) \,,\,%
{\displaystyle{2\pi \sigma  \over 3\lambda }}%
\leqslant -1 \\ 
\\ 
{\displaystyle{-3 \over 4\pi }}%
W_{0}\left( 
{\displaystyle{2\pi \sigma  \over 3\lambda }}%
\exp \left( 
{\displaystyle{2\pi \sigma  \over 3\lambda }}%
\right) \right) \,,\,%
{\displaystyle{2\pi \sigma  \over 3\lambda }}%
>-1
\end{array}
\right.
\end{equation}
is the solution which coincides with $\xi _{c}$. It gives the leading term
of $\left\langle \phi \right\rangle $ in the limit $\lambda \rightarrow 0$.
The other solution is: 
\begin{equation}
\xi ^{2}=\left\{ 
\begin{array}{c}
{\displaystyle{-3 \over 4\pi }}%
W_{0}\left( 
{\displaystyle{2\pi \sigma  \over 3\lambda }}%
\exp \left( 
{\displaystyle{2\pi \sigma  \over 3\lambda }}%
\right) \right) \,,\,%
{\displaystyle{2\pi \sigma  \over 3\lambda }}%
\leqslant -1 \\ 
\\ 
{\displaystyle{-3 \over 4\pi }}%
W_{-1}\left( 
{\displaystyle{2\pi \sigma  \over 3\lambda }}%
\exp \left( 
{\displaystyle{2\pi \sigma  \over 3\lambda }}%
\right) \right) \,,\,%
{\displaystyle{2\pi \sigma  \over 3\lambda }}%
>-1
\end{array}
\right.
\end{equation}
which gives the leading term of $\left\langle \phi \right\rangle $ in the
limit $\lambda \rightarrow \infty $. Moreover, even when the classical
picture is lost, $\sigma >0$, in which case the potential does show only one
minima, there is a solution which shows symmetry breaking in the limit $%
\lambda \rightarrow \infty $: 
\begin{equation}
\xi ^{2}=%
{\displaystyle{-3 \over 4\pi }}%
W_{-1}\left( -%
{\displaystyle{\pi m_{0}^{2} \over 6\lambda }}%
\exp \left( 
{\displaystyle{2\pi \sigma  \over 3\lambda }}%
\right) \right) \text{.}
\end{equation}
The nice behavior of these solutions relies on the fact that, for $\lambda
\rightarrow 0,$ the first solution keep the mass constant (while the
coupling constants goes to zero) and, for $\lambda \rightarrow \infty $, the
second solution gives a very large mass such that the ratio between the
coupling constants and mass goes to zero. To see this, let us calculate the
expression of the potential after we apply the Gaussian transformations. It
follows from the analysis of the last sections that the interacting measure
is equal to (up to our boundary conditions): 
\begin{equation}
d\mu _{\Lambda }=Z_{\Lambda }^{-1}e^{-\int_{x\in \Lambda }:\lambda \phi
\left( x\right) ^{4}+\sigma \phi \left( x\right) ^{2}-%
{\textstyle{m^{2} \over 2}}%
\left( \phi \left( x\right) -\xi \right) ^{2}:_{m_{0}}}d\mu _{C_{m},\xi }
\end{equation}
and after the change of the normal ordering in the exponent, with the values
of $m$ and $\xi $ given by the self-consistency equations, the coefficients
of :$\phi ^{2}$: and :$\phi $: cancel out, the result being: 
\begin{equation}
d\mu _{\Lambda }=Z_{\Lambda }^{-1}e^{-\int_{x\in \Lambda }:\lambda \phi
\left( x\right) ^{4}+4\lambda \xi \phi \left( x\right) ^{3}:_{m,\xi }}d\mu
_{C_{m},\xi }\text{.}
\end{equation}
Finally we rescale the mass at the unity by using the rescaling identity 
\cite{6}: 
\begin{equation}
d\mu _{\Lambda ^{\prime }}=Z_{\Lambda ^{\prime }}^{-1}e^{-\int_{x\in \Lambda
^{\prime }}:%
{\textstyle{1 \over 8\xi ^{2}}}%
\phi \left( x\right) ^{4}+%
{\textstyle{1 \over 2\xi }}%
\phi \left( x\right) ^{3}:_{1,\xi }}d\mu _{C_{1},\xi }\text{,}  \label{int}
\end{equation}
with $\Lambda ^{\prime }=\Lambda /m^{2}$. Note that we have omitted the free
term of the potential, $\varepsilon \left( m,\xi \right) $, which only
shifts the energies and is unimportant for the expectations values. Thus one
can see that the small coupling constant regime is achieved for large values
of $\xi $. However, even for large values of $\xi $, the convergence of the
thermodynamic limit cannot be proved by an ordinary cluster expansion. This
is due to the fact that the potential $V=%
{\textstyle{1 \over 8\xi ^{2}}}%
\phi ^{4}+%
{\textstyle{1 \over 2\xi }}%
\phi ^{3}$ is not bounded from below uniformly in $\xi $. When the field is
localized near to $-\xi $, it behaves as $-\xi ^{2}$. Nevertheless, an
expansion in phase boundaries $\left[ \text{1}\right] $ will solve the
problem. We can give now the asymptotic expansion of Schwinger functions. As
usual, we will apply the formula for integration by parts to the interaction 
\ref{int}: 
\begin{equation}
\textstyle\int%
\text{:}\phi \left( x\right) ^{m}\text{:}R\left( \phi \right) d\mu _{\Lambda
}=%
\textstyle\int%
\textstyle\int%
C_{1}\left( x-y\right) \text{:}\phi \left( x\right) ^{m-1}\text{:}\left\{ 
{\displaystyle{\delta R \over \delta \phi \left( y\right) }}%
-R%
{\displaystyle{\delta V \over \delta \phi \left( y\right) }}%
\right\} dyd\mu _{\Lambda }\text{.}
\end{equation}
The formula is still true for the case when the normal ordering is with
respect to a Gaussian measure with a nonzero mean value (see Appendix). For $%
\left\langle \phi \right\rangle $ we have to apply this formula three times
to find the first correction to $\xi $. Redenoting $\Lambda ^{\prime }$ as $%
\Lambda $, it follows: 
\begin{equation}
\begin{array}{l}
\left\langle \phi \left( x\right) \right\rangle =\int \phi \left( x\right)
d\mu _{\Lambda }=%
\textstyle\int%
(:\phi \left( x\right) :_{1,\xi }+\xi )d\mu _{\Lambda } \\ 
\\ 
=\xi +\frac{3}{2\xi ^{3}}\left[ \int C_{1}\left( x\right) ^{3}dx-\frac{9}{2}%
\int C_{1}\left( x\right) C_{1}\left( y\right) C_{1}\left( x-y\right)
^{2}dxdy\right] +o\left( 1/\xi ^{4}\right) \text{.}
\end{array}
\end{equation}
For the connected part of the two point Schwinger function, it follows: 
\begin{equation}
\begin{array}{c}
\begin{array}{c}
\left\langle \phi \left( x\right) \phi \left( y\right) \right\rangle
-\left\langle \phi \left( x\right) \right\rangle \left\langle \phi \left(
y\right) \right\rangle =\int :\phi \left( x\right) :_{1,\xi }:\phi \left(
y\right) :_{1,\xi }d\mu _{\Lambda }+o\left( 1/\xi ^{3}\right) \\ 
\\ 
=C_{1}\left( x-y\right) +\frac{9}{2\xi ^{2}}\int \int C_{1}\left( x-z\right)
C_{1}\left( y-u\right) C_{1}\left( z-u\right) ^{2}dzdu+o\left( 1/\xi
^{3}\right) \text{.}
\end{array}
\end{array}
\end{equation}
If we go back to the original scale, we can conclude: 
\begin{equation}
\left\{ 
\begin{array}{c}
\left\langle \phi \left( x\right) \right\rangle =\xi +%
{\displaystyle{0.021 \over \xi ^{3}}}%
+o\left( 1/\xi ^{4}\right) \\ 
\\ 
\left\langle \phi \left( x\right) \phi \left( y\right) \right\rangle
_{T}=C_{m}\left( x-y\right) +%
{\displaystyle{5.6\times 10^{-4} \over \xi ^{2}}}%
+o\left( 1/\xi ^{3}\right)
\end{array}
\right. \text{.}
\end{equation}

\section{Conclusions}

\noindent There are two important consequences of the analysis. First, it
was unclear before what is the meaning of the parameter $m$. Second, it
follows from the last section that this approximation is very precise for
certain range of coupling constants. In that regime, it is easyly to see
that $m$ is in fact an approximation of the self-interacting field mass,
i.e. the singular eigenvalue of the mass operator: $\hat{M}=\sqrt{\vec{P}^{2}%
}$. About the $\lambda \phi ^{4}+\sigma \phi ^{2}$ model, one can see that
it is completely determined by only one parameter, the mean value of the
field, which is experimentally measurable.

\begin{acknowledgement}
The author gratefully acknowledges support, under the direction of J.
Miller, by the State of Texas through the Texas Center for Superconductivity
and the Texas Higher Education Coordinating Board Advanced Technology
Program, and by the Robert A. Welch Foundation.
\end{acknowledgement}

\section{Appendix}

\noindent We prove first that the Wick powers are orthogonally if and only
if the measure is Gaussian i.e.: $\left\langle e^{a\phi }\right\rangle =\exp %
\left[ 
{\textstyle{a^{2} \over 2}}%
\left\langle :\phi :^{2}\right\rangle +a\left\langle \phi \right\rangle %
\right] $. This can be seen from: 
\begin{equation}
{\displaystyle{d \over da}}%
\left\langle e^{a\phi }\right\rangle =\left\langle \phi e^{a\phi
}\right\rangle =\left\langle :\phi :e^{a\phi }\right\rangle +\left\langle
\phi \right\rangle \left\langle \phi e^{a\phi }\right\rangle =\left\langle
:\phi ::e^{a\phi }:\right\rangle \left\langle e^{a\phi }\right\rangle
+\left\langle \phi \right\rangle \left\langle e^{a\phi }\right\rangle
\end{equation}
Using the orthogonality of the Wick powers we can continue 
\begin{equation}
{\displaystyle{d \over da}}%
\left\langle e^{a\phi }\right\rangle =a\left\langle :\phi :^{2}\right\rangle
\left\langle e^{a\phi }\right\rangle +\left\langle \phi \right\rangle
\left\langle e^{a\phi }\right\rangle
\end{equation}
equation which integrated out (together with the initial condition $\left.
\left\langle e^{a\phi }\right\rangle \right| _{a=0}=1$) leads to the desired
expression. The other implication follows from 
\begin{equation}
:\exp \left( a\phi \right) ::\exp \left( b\varphi \right) :=%
{\displaystyle{\exp \left( a\phi +b\varphi \right)  \over \left\langle \exp a\phi \right\rangle \left\langle \exp b\varphi \right\rangle }}%
=:\exp \left( a\phi +b\varphi \right) :%
{\displaystyle{\left\langle \exp \left( a\phi +b\varphi \right) \right\rangle  \over \left\langle \exp a\phi \right\rangle \left\langle \exp b\varphi \right\rangle }}%
\end{equation}
so, for $\phi $ and $\varphi $ jointly Gaussian variables (with mean
different by zero): 
\begin{equation}
\begin{tabular}{l}
$\left\langle :\exp \left( a\phi \right) ::\exp \left( b\varphi \right)
:\right\rangle $ \\ 
\\ 
$=%
{\displaystyle{\exp \left( %
{\textstyle{1 \over 2}}\left\langle :a\phi +b\varphi :^{2}\right\rangle +\left\langle a\phi +b\varphi \right\rangle \right)  \over \exp \left[ %
{\textstyle{a^{2} \over 2}}\left\langle :\phi :^{2}\right\rangle +a\left\langle \phi \right\rangle \right] \exp \left[ %
{\textstyle{b^{2} \over 2}}\left\langle :\varphi :^{2}\right\rangle +b\left\langle \varphi \right\rangle \right] }}%
=\exp \left( ab\left\langle :\phi ::\varphi :\right\rangle \right) $%
\end{tabular}
\end{equation}
Taking derivatives in respect to $a$ and $b$ and then the limit $a,$ $%
b\rightarrow 0$, we can form any Wick power of $\phi $ and $\varphi $.
Because of the expression on the right side, it is obvious that the Wick
powers are orthogonally.

\noindent {\bf Integration by parts} (first formulation) 
\begin{equation}
\begin{tabular}{l}
$\int :\phi \left( f\right) :_{C,\xi }R\left( \phi \right) d\mu _{C,\xi
}=\int f\left( x\right) C\left( x-y\right) \delta R/\delta \phi \left(
y\right) dxdyd\mu _{C,\xi }$.
\end{tabular}
\end{equation}
It is enough to prove this formula for the case: $R\left( \phi \right) =\exp 
\left[ i\phi \left( g\right) \right] $. In this case: 
\begin{equation}
\begin{tabular}{l}
$\int :\phi \left( f\right) :_{C,\xi }e^{i\phi \left( g\right) }d\mu _{C,\xi
}=-i%
{\displaystyle{d \over d\lambda }}%
\int :e^{i\lambda \phi \left( f\right) }:_{C,\xi }e^{i\phi \left( g\right)
}d\mu _{C,\xi }$ \\ 
\\ 
$=-i%
{\displaystyle{d \over d\lambda }}%
\int 
{\displaystyle{e^{i\phi \left( g+\lambda f\right) } \over \left\langle e^{i\phi \left( \lambda f\right) }\right\rangle }}%
d\mu _{C,\xi }=-i%
{\displaystyle{d \over d\lambda }}%
{\displaystyle{\exp \left[ -\frac{1}{2}C\left( g+\lambda f,g+\lambda f\right) +i\left( \xi ,g+\lambda f\right) \right]  \over \exp \left[ -\frac{1}{2}C\left( \lambda f,\lambda f\right) +i\left( \xi ,\lambda f\right) \right] }}%
$ \\ 
\\ 
$=iC\left( f,g\right) \exp \left[ -\frac{1}{2}C\left( g,g\right) +i\left(
\xi ,g\right) \right] =i\int f\left( x\right) C\left( x-y\right) g\left(
y\right) dxdy\left\langle e^{i\phi \left( g\right) }\right\rangle $ \\ 
\\ 
$=\int f\left( x\right) C\left( x-y\right) 
{\displaystyle{\delta \exp \left[ i\phi \left( g\right) \right]  \over \delta \phi \left( y\right) }}%
dxdyd\mu _{C,\xi }$,
\end{tabular}
\end{equation}
where $\left( \cdot ,\cdot \right) $ is the scalar product in $L_{2}\left( 
{\Bbb R}^{2}\right) $ and $f$ and $g$ were considered in $L_{1}\left( {\Bbb R%
}^{2}\right) $. Next we show that: $\delta :\phi \left( f\right)
^{n}:/\delta \phi \left( x\right) =:\delta \phi \left( f\right) ^{n}/\delta
\phi \left( x\right) :$, where the normal ordering is with respect to an
arbitrary measure, which is helpful when one computes $\delta R/\delta \phi $%
. Indeed: 
\begin{equation}
\begin{tabular}{l}
$%
{\displaystyle{\delta  \over \delta \phi \left( x\right) }}%
:\phi \left( f\right) ^{n}:=%
{\displaystyle{\delta  \over \delta \phi \left( x\right) }}%
\left. 
{\displaystyle{d^{n} \over d\lambda ^{n}}}%
:e^{\lambda \phi \left( f\right) }:\right| _{\lambda =0}=%
{\displaystyle{\delta  \over \delta \phi \left( x\right) }}%
\left. 
{\displaystyle{d^{n} \over d\lambda ^{n}}}%
{\displaystyle{e^{\lambda \phi \left( f\right) } \over \left\langle e^{\lambda \phi \left( f\right) }\right\rangle }}%
\right| _{\lambda -0}$ \\ 
\\ 
$=\left. 
{\displaystyle{d^{n} \over d\lambda ^{n}}}%
{\displaystyle{\lambda f\left( x\right) e^{\lambda \phi \left( f\right) } \over \left\langle e^{\lambda \phi \left( f\right) }\right\rangle }}%
\right| _{\lambda =0}=nf\left( x\right) \left. 
{\displaystyle{d^{n-1} \over d\lambda ^{n-1}}}%
{\displaystyle{e^{\lambda \phi \left( f\right) } \over \left\langle e^{\lambda \phi \left( f\right) }\right\rangle }}%
\right| _{\lambda =0}=nf\left( x\right) :\phi \left( f\right) ^{n-1}:$ \\ 
\\ 
$=:%
{\displaystyle{\delta  \over \delta \phi \left( x\right) }}%
\phi \left( f\right) ^{n}:$.
\end{tabular}
\end{equation}
{\bf Integration by parts} (second formulation) 
\begin{equation}
\begin{tabular}{l}
$\int :\phi \left( f\right) ^{n}:_{C,\xi }R\left( \phi \right) d\mu _{C,\xi
}=\int f\left( x\right) C\left( x-y\right) :\phi \left( f\right)
^{n-1}:_{1,\xi }\delta R/\delta \phi \left( y\right) dxdyd\mu _{C,\xi }$.
\end{tabular}
\end{equation}
We start with the identity: 
\begin{equation}
:\phi \left( f\right) ^{n}:_{C,\xi }=:\phi \left( f\right) :_{C,\xi }:\phi
\left( f\right) ^{n-1}:_{C,\xi }-\int f\left( x\right) C\left( x-y\right) :%
\frac{\delta \phi \left( f\right) ^{n-1}}{\delta \phi \left( y\right) }%
:_{C,\xi }dxdy\text{,}
\end{equation}
which can be proven as follow: 
\begin{equation}
\begin{tabular}{l}
$:\phi \left( f\right) ^{n}:_{C,\xi }=\left. 
{\displaystyle{d^{n} \over d\lambda ^{n}}}%
:e^{\lambda \phi \left( f\right) }:_{1,\xi }\right| _{\lambda =0}=\left. 
{\displaystyle{d^{n} \over d\lambda ^{n}}}%
{\displaystyle{\exp \left[ \lambda \phi \left( f\right) \right]  \over \left\langle \exp \left[ \lambda \phi \left( f\right) \right] \right\rangle _{C,\xi }}}%
\right| _{\lambda =0}$ \\ 
\\ 
$=\left. 
{\displaystyle{d^{n} \over d\lambda ^{n}}}%
\exp \left[ \lambda \phi \left( f\right) -\frac{1}{2}C\left( f,f\right)
-\lambda \left( \xi ,f\right) \right] \right| _{\lambda =0}$ \\ 
\\ 
$=\left. 
{\displaystyle{d^{n-1} \over d\lambda ^{n-1}}}%
\left[ \left[ \phi \left( f\right) -\left( \xi ,f\right) \right] 
{\displaystyle{\exp \left[ \lambda \phi \left( f\right) \right]  \over \left\langle \exp \left[ \lambda \phi \left( f\right) \right] \right\rangle _{C,\xi }}}%
-\lambda C\left( f,f\right) 
{\displaystyle{\exp \left[ \lambda \phi \left( f\right) \right]  \over \left\langle \exp \left[ \lambda \phi \left( f\right) \right] \right\rangle _{C,\xi }}}%
\right] \right| _{\lambda =0}$ \\ 
\\ 
$=:\phi \left( f\right) :_{C,\xi }:\phi \left( f\right) ^{n-1}:_{C,\xi
}-\left. 
{\displaystyle{d^{n-1} \over d\lambda ^{n-1}}}%
\int f\left( x\right) C\left( x-y\right) 
{\displaystyle{\delta  \over \delta \phi \left( y\right) }}%
{\displaystyle{\exp \left[ \lambda \phi \left( f\right) \right]  \over \left\langle \exp \left[ \lambda \phi \left( f\right) \right] \right\rangle _{C,\xi }}}%
\right| _{\lambda =0}dxdy$ \\ 
\\ 
$=:\phi \left( f\right) :_{C,\xi }:\phi \left( f\right) ^{n-1}:_{C,\xi
}-\int f\left( x\right) C\left( x-y\right) 
{\displaystyle{\delta  \over \delta \phi \left( y\right) }}%
:\phi \left( f\right) ^{n-1}:_{C,\xi }dxdy$.
\end{tabular}
\end{equation}
Then we can continue: 
\begin{equation}
\begin{tabular}{l}
$\int :\phi \left( f\right) ^{n}:_{C,\xi }R\left( \phi \right) d\mu _{C,\xi
}=%
\textstyle\int%
:\phi \left( f\right) :_{C,\xi }:\phi \left( f\right) ^{n-1}:_{C,\xi
}R\left( \phi \right) d\mu _{C,\xi }$ \\ 
\\ 
$-\int \int \int f\left( x\right) C\left( x-y\right) 
{\displaystyle{\delta  \over \delta \phi \left( y\right) }}%
:\phi \left( f\right) ^{n-1}:_{C,\xi }R\left( \phi \right) dxdyd\mu _{C,\xi }
$ \\ 
\\ 
$=\int \int f\left( x\right) C\left( x-y\right) 
{\displaystyle{\delta  \over \delta \phi \left( y\right) }}%
\left[ :\phi \left( f\right) ^{n-1}:_{C,\xi }R\left( \phi \right) \right]
dxdyd\mu _{C,\xi }$ \\ 
\\ 
$-\int \int \int f\left( x\right) C\left( x-y\right) 
{\displaystyle{\delta  \over \delta \phi \left( y\right) }}%
:\phi \left( f\right) ^{n-1}:_{C,\xi }R\left( \phi \right) dxdyd\mu _{C,\xi }
$,
\end{tabular}
\end{equation}
which leads to the desired expression.

\end{document}